\documentclass[3p,times]{elsarticle}

\usepackage{amsmath,amssymb,amsfonts}
\PassOptionsToPackage{hyphens}{url}\usepackage{hyperref}
\usepackage{amsmath,amssymb,amsfonts}
\usepackage{algorithmic}
\usepackage{graphicx}
\usepackage{textcomp}
\usepackage{xcolor}
\usepackage{caption}
\usepackage{subcaption}
\usepackage{booktabs}
\usepackage{float}
\usepackage{amsmath}
\usepackage{booktabs}
\usepackage{fancyhdr}
\usepackage{booktabs}
\usepackage{pifont}
\usepackage{tikz}
\usepackage{filecontents}
\usepackage[labelfont=bf]{caption}
\usepackage{breakcites}
\usepackage{lineno}
\usepackage[linesnumbered,ruled,vlined,onelanguage]{algorithm2e}
\usepackage{hyphenat}
\usepackage{subcaption}
\usepackage{etoolbox}
\usepackage{verbatim}
\usepackage{pdfpages}

\modulolinenumbers[5]

\begin{document}
\begin{sloppypar}

\begin{frontmatter}

% Siamak: my sugesstion:
\title{XG-BoT: An Explainable Deep Graph Neural Network \\for Botnet Detection and Forensics} 

% \title{Anomal-E: A Self-Supervised Graph Neural Network based Network Intrusion Detection System}
\cortext[cor1]{Accepted by Internet of Things, Elsevier}
\cortext[cor]{Corresponding author}


\author[1]{Wai Weng Lo\corref{cor}}
\ead{w.w.lo@uq.net.au}

\author[1]{Gayan Kulatilleke}
\ead{g.kulatilleke@uq.net.au}

\author[1]{Mohanad Sarhan}
\ead{m.sarhan@uq.net.au}

\author[1]{Siamak Layeghy}
\ead{siamak.layeghy@uq.net.au}

\author[1]{Marius Portmann}
\ead{marius@itee.uq.edu.au}

\address[1]{School of ITEE, The University of Queensland, Brisbane, Australia}
\date{}

\begin{abstract}
In this paper, we propose XG-BoT, an explainable deep graph neural network model for botnet node detection. The proposed model comprises a botnet detector and an explainer for automatic forensics. The XG-BoT detector can effectively detect malicious botnet nodes in large-scale networks. Specifically, it utilizes a grouped reversible residual connection with a graph isomorphism network to learn expressive node representations from botnet communication graphs. The explainer, based on the GNNExplainer and saliency map in XG-BoT, can perform automatic network forensics by highlighting suspicious network flows and related botnet nodes. We evaluated XG-BoT using real-world, large-scale botnet network graph datasets. Overall, XG-BoT outperforms state-of-the-art approaches in terms of key evaluation metrics. Additionally, we demonstrate that the XG-BoT explainers can generate useful explanations for automatic network forensics.
\end{abstract}

\begin{keyword}
Graph neural network, Graph representation learning, Botnet detection, Digital forensics, Anomaly detection
\end{keyword}

\end{frontmatter}

\section{Introduction}

A botnet is a computer network that consists of compromised victim computers or IoT devices (bots) controlled by a "botmaster" to perform malicious activities. The victims are usually used for distributed denial of service (DDoS) attacks, phishing, and malware propagation. Due to dynamic changes in network flows (i.e., rapid changes in botnet sizes), it is very difficult to detect botnet nodes effectively. Current machine learning (ML)-based botnet detection approaches \cite{gu2008botminer,bilge2012disclosure} require a huge amount of domain knowledge and manual effort from experts for the extraction of features such as packet sizes, packet byes, and corresponding protocols for feature extraction. Furthermore, network traffic can be encrypted and intentionally manipulated (i.e., Payload mutation) \cite{cheng2011evasion} to evade ML-based NIDS.

The botnet detection problem can be solved by a graph-based approach, where network flows are represented as communication flows, and nodes are mapped as the victims and attackers. This makes it possible to consider the overall graph patterns, in addition to the network flows and features, for botnet detection. Previous works \cite{chowdhury2017botnet,abou2019graph} only used the graph topological pattern of the botnet network graphs for detection, ignoring network flow features. These works considered graph centrality as a feature for botnet detection, which might not describe the corresponding botnet patterns sufficiently. 

\begin{figure*}[!t]
    \centering
        \includegraphics[width=0.65\columnwidth]{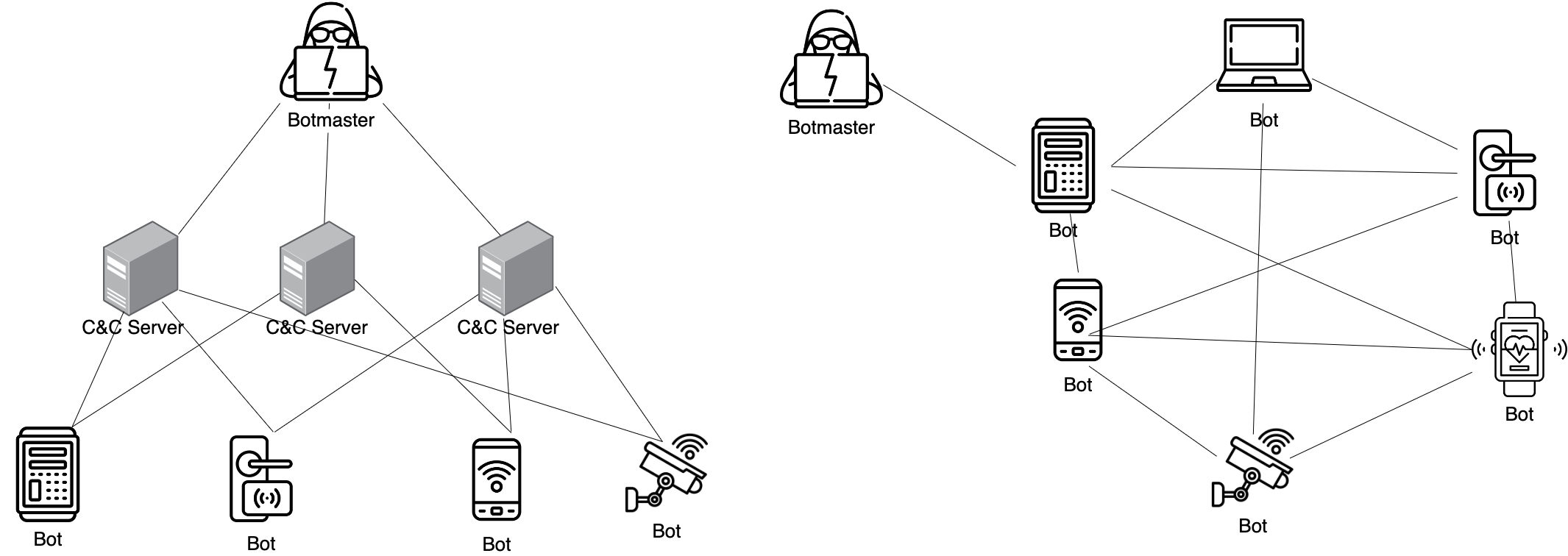}
        %\label{rfidtest_yaxis}
    \caption{Botnet Overview \textbf{Left}: C2 Botnet \textbf{Right:} P2P Botnet}
    \label{fig:bbotnet}
\end{figure*}

Graph Neural Networks (GNNs) \cite{wu2020comprehensive} represent a fast-growing field in machine learning. They can automatically capture the graph data structure to perform downstream tasks, such as node classification. Since the botnet network graph consists of rich structural information, the corresponding graph structures can be utilized for automatic botnet detection based on the GNN. Thus, most recent works~\cite{zhou2020automating,zhang2021practical} have explored the use of vanilla graph convolutional networks (GCNs) for botnet node detection by converting the problem into a node classification problem. However, vanilla GCNs are susceptible to over-smoothing, vanishing gradient, and over-fitting problems \cite{li2019deepgcns}, and the model performance can degrade significantly with an increase in the number of graph convolutional layers. 

In this paper, we propose a deep GNN model to against sophisticated botnet attacks. We theorize that GNN models need to be deeper to capture some of the hidden topological botnet patterns. To achieve this task, we first used real-world botnet graph datasets, a decentralized botnet P2P, and a centralized botnet C2 dataset \cite{zhou2020automating}, as illustrated in Fig. \ref{fig:bbotnet}, and trained the GNN model to learn the botnet topological patterns. The major problem is that most deep GNN approaches suffer from the over-smoothing problem \cite{li2019deepgcns}. Botnet topological patterns can come in drastically varied depths \cite{vormayr2017botnet}, and deeper models are required to detect some of the hidden topological patterns. Thus, we propose XG-BoT, which is based on grouped reversible residual connections~\cite{li2021training}, which can effectively handle model performance degradation (i.e., vanishing gradient problem) caused by a higher number of layers with a powerful graph isomorphism network (GIN)~\cite{xu2018powerful} for effective botnet detection. 

Flagging a malicious botnet node is not a trivial process. False alarms can distract the network administrator and increase the level of irrelevant workload. Extensive human resources are required to review the model's detection results, which is inefficient and costly. Therefore, in this paper, we also investigate explainability approaches, specifically the GNNExplainer \cite{ying2019gnnexplainer} and saliency maps \cite{kasanishi2021edge}, to provide intuitive explanations for model predictions based on highlighting suspicious network flows and related botnet nodes. 

Overall, the aim of this study is to propose a novel and explainable deep GNN-based botnet detection system. Most related works, such as \cite{mcdermott2018botnet, antonakakis2017understanding, ahmed2020deep}, consider network flows independently, without taking into account their interconnected relationship, which is important in botnet detection. On the other hand, other GNN approaches, such as those presented in \cite{zhou2020automating} and \cite{zhang2021practical}, are susceptible to over-smoothing problems and lack explainability for network forensics. To address these limitations, we present the XG-BoT model, which captures graph patterns for sophisticated botnet detection. Additionally, we utilize GNNExplainer and saliency maps to highlight suspicious network flows and botnet nodes, which makes the detection process more transparent and interpretable for automatic network forensics, and those features are lacking in the current botnet detection related works. Our results demonstrate that the proposed approach achieves state-of-the-art performance in terms of the detection rate. It also can generate useful explanations based on subgraph visualisation for automatic forensics, which indicates the potential for utilising deep GNN approaches in automatic botnet detection and forensics.

In summary, the key contributions of this paper are as follows:

\begin{itemize}
\setlength\itemsep{0.8em}
\item We propose and implement XG-BoT, a deep, explainable, graph neural network-based model that can detect botnets within large-scale communication networks. XG-BoT can also enable automatic forensics by highlighting suspicious network flows and related botnet nodes using GNNExplainer and saliency maps. This, until now, has been lacking in current botnet detection approaches.

% We propose XG-BoT, a deep graph neural network-based model that can detect botnets within large-scale communication networks while providing explanations for its predictions. XG-BoT also includes a feature for automatic forensics, using GNNExplainer and saliency maps to highlight suspicious network flows and related botnet nodes. To the best of our knowledge, this feature has not been included in previous botnet detection-related works.

%Please check intended meaning is retained.

\item The comprehensive evaluation of the proposed XG-BoT approach, using two datasets, indicates that it can achieve superior performance compared to state-of-the-art approaches. Additionally, the experiments demonstrate that useful explainable results can be generated for automatic network forensics.
\end{itemize}

The paper is organized as follows: In Section~\ref{related_works}, we introduce the latest works related to botnet detection. In Section~\ref{method}, we describe our proposed method. Section~\ref{chap:exp} describes the experimental settings. In Section~\ref{results}, we provide an evaluation of the model. Section~\ref{explainable_results} discusses explainability, and Section~\ref{summary} summarises the paper.
% The paper is organised as follows: In Section~\ref{related_works}, we introduce the latest works related to botnet detection. In Section~\ref{method}, we describe our proposed method. Section~\ref{chap:exp} describes the experimental settings. In Section~\ref{results}, we provide an evaluation of the model. Section~\ref{explainable_results} discusses explainability, and Section~\ref{summary} summarises the paper.

\section{Botnets}

A botnet is a collection of bots, which are computers or IoT devices that have been compromised through malicious software. Bots can be controlled remotely by attackers and are connected to botmasters to receive commands to perform malicious activities. Botmasters are attackers who control botnets by issuing commands to the bot to perform malicious actions. Botmasters typically remain hidden from the public and evade detection by law enforcement. One of the most common botnet architectures is the centralized command-and-control infrastructure (C2) \cite{vormayr2017botnet}, as shown on the left of Fig \ref{fig:bbotnet}, consisting of bots and a centralized control entity. In C2, the botmasters attempt to use one or more network protocols to command victim computers and coordinate their actions. The instructions can range from the execution of denial-of-service attacks to malicious software propagation. The centralized approach is very similar to the traditional client and server architecture. It can be implemented through the internet relay chatbot (IRC) protocol, where all bots establish a communication channel with the botmaster. 

The C2 architecture enables easy and direct communication with the bot and allows the botmaster to monitor the global distribution and bot status. However, the main problem with the C2 architecture is its single point of failure, which allows law enforcement to shut down the botnet easily. This provides motivation for the use of a decentralized P2P \cite{vormayr2017botnet} botnet architecture, as shown on the right of Fig \ref{fig:bbotnet}. Due to the decentralized nature of the C2 architecture, P2P Botnets are more resilient and more difficult to shut down because there is no central C2 server that can be disabled.

\section{Related Works} \label{related_works}

McDermott et al.~\cite{mcdermott2018botnet} used a deep learning approach to perform botnet detection on IoT networks based on a bidirectional Long Short-Term Memory Recurrent Neural Network (BLSTM-RNN) model to detect botnet activity within consumer IoT devices and networks. They first used word embeddings to encode and convert packets into a tokenized integer format for feature extraction. Then, the word embedding vector was fed to the BLSTM-RNN model to detect botnet activity. The authors applied four attack vectors used by the Mirai botnet malware \cite{antonakakis2017understanding} for the performance evaluation. Shi et al.~\cite{shi2020deepbot} proposed a deep learning method to detect and classify botnets on the extracted features from the input network traffic by using LSTM and RNN models. Ahmed et al.~\cite{ahmed2020deep} proposed a deep feed-forward neural network model for botnet detection and compared the proposed method with the Support Vector Machine and Naive Bayes algorithms. Kozik et al. \cite{kozik2018scalable} developed an attack detection platform for IoT applications based on Extreme Learning Machines (ELM) and the Apache Spark framework. They first used the CTU-13 Netflow dataset collected from an IoT network and then applied the proposed platform for botnet detection. The results show that the approach achieves high accuracy values of 0.99, 0.76, and 0.95 for scanning, C2 and infected host scenarios, respectively.

In \cite{pektacs2019deep}, the authors proposed a botnet detector by combining both network flow and graph pattern information. They first built the network graph to represent network connections between hosts and then extracted statistical-based information from the graph features and network flow for feature extraction to train the 1D CNN-RNN to identify botnets. The CTU-13 and ISOT datasets were used to evaluate the model's effectiveness. The experimental results show that an overall accuracy of 99.3\%  and an F1-score of 99.1\% were achieved. In \cite{moodi2019new}, the authors created a 28 Standard Android Botnet Dataset (28-SABD) and an Android botnet malware dataset, including 14 families of Android botnet malware traffic. They used the ensemble K-Nearest Neighbors (KNN) technique to improve the overall detection accuracy. However, the proposed method only obtained an overall accuracy of 94\%, which indicates that further improvement of the proposed method is needed. In~\cite{al2020unsupervised}, the authors introduced an unsupervised botnet detection method for IoT based on the One-Class Support Vector Machine (OCSVM). They applied Grey Wolf optimization (GWO) to optimize the hyperparameters of the OCSVM to improve the botnet detection performance.

Graph-based machine learning methods for botnet detection are explored in \cite{chowdhury2017botnet,abou2019graph}. The authors designed a graph-based technique to detect botnet attacks. They first built a botnet communication graph by representing hosts as nodes and edges as network flows between them. Subsequently, they extracted the graph centrality features, such as the degree centrality, for feature extraction and applied various machine learning algorithms to detect the botnets. However, these works considered graph centrality as a feature, and this might not be able to completely capture the hidden botnet topological patterns sufficiently.  

In~\cite{zhou2020automating,zhang2021practical}, the authors used a basic 'vanilla' GCN to detect botnets. They exploited generated botnet traffic and normal traffic created by using the CTU-13 ~\cite{garcia2014empirical} dataset to create botnet communication graphs and applied a 12-layer GCN for botnet node classification. Botnet communication graphs do not contain any node features for botnet detection, and those approaches rely on the graph pattern structure of the communication graph for botnet node classification. Nevertheless, these approaches suffer from the over-smoothing problem \cite{liu2020towards}. It has been indicated that, as the number of graph convolutional layers increases, all node embeddings over a graph will converge to indistinguishable vectors, which can lower the performance in downstream tasks~\cite{li2018deeper}. Zhou et al.~\cite{zhou2020automating} theorized that the GNN model needs to reach a certain depth in order to capture some of the hidden topological botnet patterns. Thus, further investigation of the application of deep GNNs for botnet detection is critical. 

\begin{table}[htbp]
\centering
\caption{Comparison of state-of-the-art botnet detection methods}
\resizebox{.9\textwidth}{!}{
\begin{tabular}{llllll}
 \toprule
  \textbf{Reference}  & \textbf{Method} & \textbf{Features} & \textbf{Datasets}  & \textbf{Support for Automatic Network Forensics}  \\
\midrule
Chowdhury et al. \cite{chowdhury2017botnet}  & Graph-based & Graph centrality & CTU-13 \cite{garcia2014empirical} &  No\\
McDermott et al. \cite{mcdermott2018botnet}  & BLSTM-RNN & Word embeddings & Custom dataset & No \\
Kozik et al. \cite{kozik2018scalable}  & ELM and Spark & Network flow features & CTU-13 \cite{garcia2014empirical} & No\\
Moodi et al. \cite{moodi2019new}  & Ensemble KNN &Network flow features & Custom dataset & No\\
Abou-Rjeili et al. \cite{abou2019graph}  & Graph-based & Graph centrality & CTU-13 \cite{garcia2014empirical} & No \\
Pektacs et al. \cite{pektacs2019deep}  & 1D CNN-RNN & Graph and network flow features & CTU-13 \cite{garcia2014empirical} and ISOT \cite{isot} & No \\
Shi et al. \cite{shi2020deepbot}  & LSTM and RNN &Network flow features & CTU-13 \cite{garcia2014empirical} & No \\
Ahmed et al. \cite{ahmed2020deep}  & Deep feed-forward NN & Network flow features & CTU-13 \cite{garcia2014empirical} & No \\
Alqahtani et al. \cite{al2020unsupervised}  & OCSVM and GWO &Network flow features & N-BaIoT \cite{meidan2018n} & No \\
Zhou et al. \cite{zhou2020automating}  & GCN & Communication graphs & CTU-13 and CAIDA \cite{zhou2020automating} & No \\
Zhang et al.~\cite{zhang2021practical}  & GCN & Communication graphs & CTU-13 and CAIDA \cite{zhou2020automating} & No \\

\textbf{XG-BoT} & \textbf{Grouped reversible GINs} & \textbf{Communication graphs} & \textbf{CTU-13 and CAIDA } \cite{zhou2020automating} & \textbf{Yes} \\

\bottomrule
\end{tabular}%
}
\label{tab:botnet_detection_comparison}%
\end{table}%

Table~\ref{tab:botnet_detection_comparison} presents a comparison of various botnet detection methods used in related studies. In contrast to these studies, our XG-BoT approach utilizes grouped reversible residual connections with GINs for botnet detection. This approach helps mitigate the over-smoothing problem present in \cite{zhou2020automating, zhang2021practical} and captures deeper hidden botnet patterns to improve detection performance. Additionally, we enable GNN explainability for automatic network forensics by highlighting highly correlated hosts and network flows, which until now has been lacking in current botnet detection approaches.

\section{Proposed Method}\label{method}
\begin{figure*}[t]
\hspace*{-1.0cm}
    \centering
        \includegraphics[width=1.15\columnwidth]{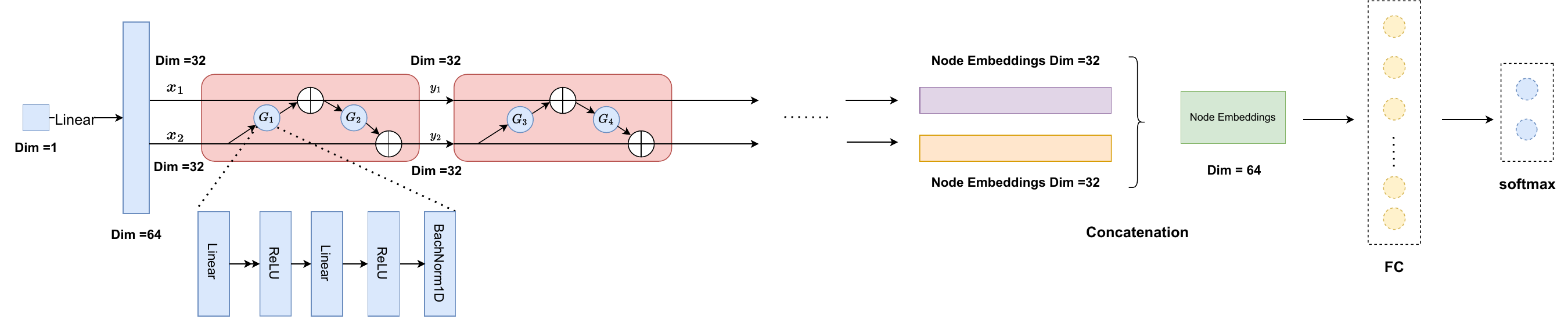}
        %\label{rfidtest_yaxis}
    \caption{Proposed of XG-BoT architecture}
    \label{fig:XG-BoT}
\end{figure*}

The proposed XG-BoT approach was designed to extract useful graph topological patterns for botnet detection in large-scale botnet graph datasets. Since botnet datasets inherit a high-class imbalance, which affects the detection performance, the proposed method aims to train very deep GCNs and capture the hidden topological patterns for botnet detection as much as possible. The goal is to improve classification performance by enhancing node representation.

The GCN utilizes a message propagation mechanism to compute node embeddings by incorporating the $K$-hop neighbours' node features. The trained GCN models can be applied to different graphs to generate embeddings for the downstream tasks (i.e., node classification). 

We consider a network $G = (N, E)$, where $N$ represents the set of hosts (vertices) and $E$ is the set of communication flows (edges). 
%$|N|$ is the number of hosts in the graph, and $|E|$ is the number of communication flows. 
The adjacency matrix $A$ is an $N \times N$ sparse matrix with $(i,j)$. Each node has a k-dimensional node feature vector, and $X \in \mathbb{R}^{N \times K}$ is a feature vector for each $N$ node. 

The $k$-th layer of a typical GCN is  
\begin{equation}
%Z^{l+1}=\sigma\left(X^{l} W_{0}^{l}+\tilde{A} X^{l} W_{1}^{l}\right)
h_v^{(k)} =   \sigma \left(W \cdot {\rm MEAN} \left\{h_u^{(k-1)},\ \forall u \in {N}(v) \cup \{v\} \right\} \right),
\end{equation}

$W^{(l)}$ is the weight matrix that will be learned for the downstream tasks. $\sigma$ is an activation function, typically ReLU, for computing node representations. Since the vanilla GCN is limited by the number of layers, we propose XG-BoT, which combines the grouped reversible residual connections~\cite{li2021training} with GINs~\cite{xu2018powerful} for botnet detection to act against the over-smoothing problem and capture deeper hidden botnet patterns. \\

\textbf{Proposed XG-BoT model:} In this study,  we utilized grouped reversible residual connections~\cite{li2021training} with GINs~\cite{xu2018powerful} to build the XG-BoT model to perform botnet node detection, as shown in Fig \ref{fig:XG-BoT}. In the XG-BoT model, the input node feature matrix $X$, which is an all one's constant vector, is transformed into a 64-dimensional vector using a linear transformation and then uniformly partitioned into $C=2$ groups across the hidden layer channel dimension. Each of the grouped GIN modules only takes the corresponding group of node features to compute the corresponding node embeddings. The forward propagation of computing embeddings $X^{\prime}$ is performed as follows:

\begin{equation}
X_{0}^{\prime}=\sum_{i=2}^{C} X_{i}
\label{eq:group_c}
\end{equation}

\begin{equation}
X_{i}^{\prime}=f_{w i}\left(X_{i-1}^{\prime}, g\right)+X_{i}, i \in\{1, \cdots, C\}
\label{eq:group_prop}
\end{equation}

\begin{equation}
X^{\prime}=X_{1}^{\prime}\| X_{2}^{\prime}
\label{eq:res}
\end{equation}

where $X_{0}^{\prime}$ represents node features split into $C$ groups for message propagation, as shown in  Equation~\ref{eq:group_c}. In Equation~\ref{eq:group_prop},  each of the GIN blocks only takes the corresponding grouped node feature  $X_{i}$ for computing node embeddings with the reversible residual connection mechanism \cite{gomez2017reversible} to minimize over-smoothing and memory consumption problems. For example, in Fig.~\ref{fig:XG-BoT}, it is assumed that there are $C=2$ groups and each of the reversible GIN blocks takes two inputs $\left(x_{1}, x_{2}\right)$ and produces two intermediate node representations $\left(y_{1}, y_{2}\right)$. The residual functions of the GIN blocks $G_{1}$ and $G_{2}$ are

\begin{equation}
\begin{aligned}
&y_{1}=x_{1}+G_{1}\left(x_{2}\right) \\
&y_{2}=x_{2}+G_{2}\left(y_{1}\right)
\end{aligned}
\end{equation}

In the final step of forward propagation, the node embeddings are reconstructed based on the concatenation operation of each of the subset node embeddings, as shown in Equation~\ref{eq:res}. To compute the final node embedding, this is fed to a fully connected (MLP) layer and Softmax layer to perform the downstream task (i.e., botnet node classification). Due to group processing, the number of training parameters decreases as the size of the group increases. This allows a deeper XG-BoT model to be built, allowing the capture of hidden topological patterns. \\

\textbf{Graph Encoder module:} As we mentioned, we adopted GIN~\cite{xu2018powerful}, a state-of-the-art graph neural network, as the GNN encoder block in Fig. \ref{fig:XG-BoT}. GIN calculates each node representation via a sum-like neighbourhood aggregation function, as shown below:
\begin{equation}
h_{v}^{(k)}=\mathrm{MLP}^{(k)}\left(\left(1+\epsilon^{(k)}\right) \cdot h_{v}^{(k-1)}+\sum_{u \in \mathcal{N}(v)} h_{u}^{(k-1)}\right)
\label{eq:gin}
\end{equation}

where $h_{v}^{(k)}$ is the node embedding of node $v_{i}$ at the $k$-th layer. The MLP isthe multi-layer perceptron. $\epsilon^{(k)}$ is either a fixed scalar or trainable parameter. We can stack $L$ layers of GIN to obtain the final node embedding  $h_{v}^{(L)}$.   Unlike vanilla GCNs \cite{kipf2016semi}, which are much less powerful than the Weisfeiler--Lehman (1-WL) \cite{shervashidze2011weisfeiler} algorithm, the sum-like aggregator in GIN can capture the structural homophily and neighbourhood homophily, which are both critical for representing the botnet behaviour patterns. In terms of the GIN encoder, we used two MLP layers with the ReLU activation function and batch normalization to extract node embeddings, as shown in Fig.~\ref{fig:XG-BoT}.

\section{Experiments} \label{chap:exp}
\subsection{Datasets}

% TTTTTTTTTTTTTTTTTTTTTTTTTTTTTTTTTTTTTTTT
\begin{table}[b]
\centering
\caption{Botnet dataset statistics for P2P}
\begin{tabular}{ c c c c c c c c}
\hline \textbf{Data Split} & \textbf{Graphs} & \textbf{Avg Nodes} & \textbf{Avg Edges} & \textbf{Avg Botnet Nodes}  %& Prediction Time (µs) 
\\   \hline

Train & 768 & 143895 & 1623217 & 3090 
\\
 Val & 96 & 143763 & 1622620 & 3093
\\
 Test & 96 & 144051 & 1624948 & 3095 
\\
\hline
\end{tabular}

\label{tab:p2p_statistic}
\end{table}

\begin{table}[t]
\centering
\caption{Botnet dataset statistics for C2}
\begin{tabular}{ c c c c c c c c}
\hline \textbf{Data Split} & \textbf{Graphs} & \textbf{Avg Nodes} & \textbf{Avg Edges} & \textbf{Avg Botnet Nodes}  %& Prediction Time (µs) 
\\   \hline

Train & 768 & 143895 & 813237 & 3211 
\\
 Val & 96 & 143763 & 812955 & 3234
\\
 Test & 96 & 144051 & 814003 & 3175 
\\
\hline
\end{tabular}

\label{tab:c2_statistic}
\end{table}

In general, a set of network flows from the datasets can be treated as graph data, as each of the network host IP addresses can be represented as a graph node, and network communication flows between each host can be represented as edges. Formally, we can define the communication graph as $ G=(V, E)$, where graphs $G$ consist of a collection $V$ of nodes and a collection $E$ of edges. An adjacency matrix format can represent the communication graphs, whereby a graph with $n$ nodes and communication flow can be represented as an adjacency matrix $A \in \mathbb{R}^{n \times n}$ with $a_{i j}=1$ if there is a communication flow between host node $i$ and host node $j$. 

We used two publicly available botnet graph datasets \cite{zhou2020automating} with P2P and C2 botnet scenarios. The botnet graph datasets were generated from the CTU-13 original NetFlow dataset \cite{garcia2014empirical}, and the botnet traffic was generated by real-world malware samples. The botnet nodes and botnet topological patterns were mixed with background traffic collected from CAIDA in 2018 to generate botnet communication graphs.

Both the P2P and C2 botnet datasets consist of 768 training graphs and 96 validation and testing graphs. Each of these graphs contains around 3,000 botnet nodes. Each node is equipped with an "all ones" constant vector as its node feature. Due to privacy concerns, the IP address of each network node was numerically relabeled to the nodes of each graph by the dataset authors. The distribution of the dataset and the statistics of the nodes and edges for each botnet graph are shown in Table~\ref{tab:p2p_statistic} and Table~\ref{tab:c2_statistic}.

\subsection{Training}
Our experiments were conducted on a virtual Linux server with a $2.3 \mathrm{GHz}$ 2-core Intel(R) Xeon(R) processor and 51 GB memory, and a Tesla P100 GPU. The proposed model was developed in  Python using several machine learning packages, such as Sckit-learn,  PyTorch Geometric, and PyTorch.

\begin{figure}[!h]
    \centering
        \includegraphics[width=0.55\columnwidth]{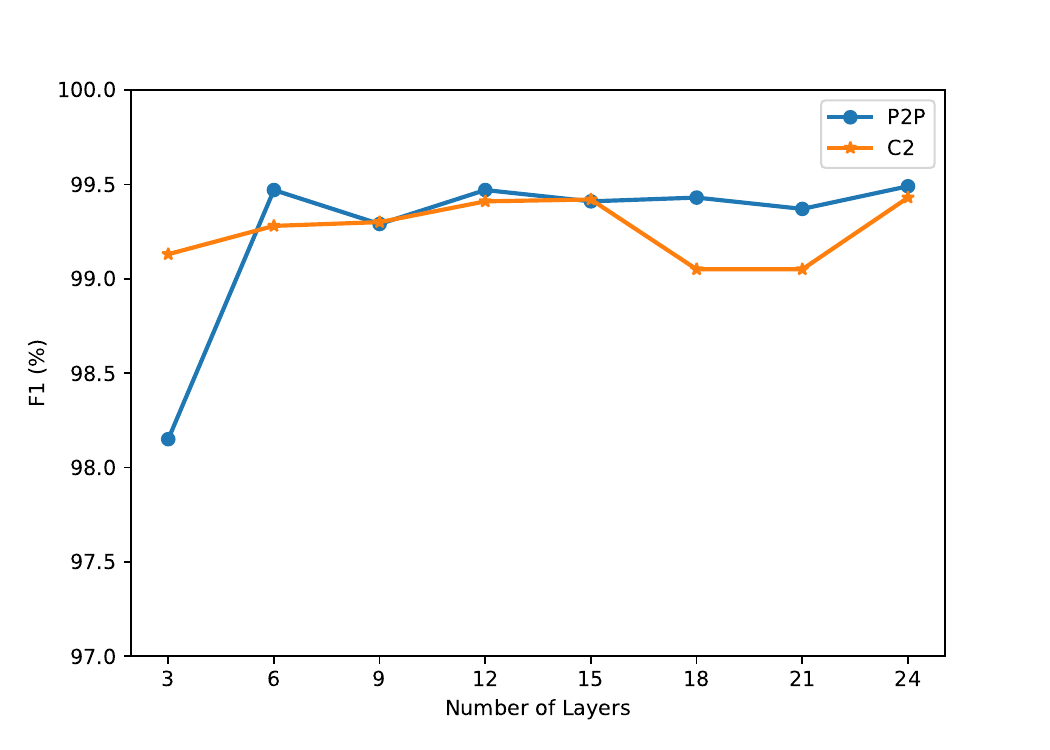}
        %\label{rfidtest_yaxis}
    \caption{Average F1-scores on P2P and C2 test sets with varying number of layers}
    \label{fig:f1_measure}
\end{figure}

For performing hyperparameter tuning, a grid search was performed to ensure the optimal settings were used. The XG-BoT grid search values are given in Table~\ref{lab:hypermanters}. Overall, we found that the optimal parameters were 15 XG-BoT layers for the C2 datasets and 6 XG-BoT layers for P2P datasets with $C=2$, 64 hidden channels (32 hidden channels for each GINs as $C=2$) and $\epsilon=0$. The results for different layers are shown in Fig \ref{fig:f1_measure}. We used the Adam optimizer with a learning rate of 0.001 to train the proposed model. 

             \begin{table}[!h]\small
            \caption{XG-BoT training time and MTTD}
            \label{anomal-e_parameters}
            \centering
            \begin{tabular}{*3l }
                \toprule
                Datasets & Training time (hrs) & Mean time to detect (MTTD) ($\mu$s)  \\
                \toprule
                C2 & 4.63 & 1.45\\
                P2P & 3.82 & 1.59\\
               \hline
            \end{tabular}
                
        \label{lab:runtime}
        \end{table}
        
%TTTTTTTTTTTTTTTTTTTTTTTTTTTTTTTTTTTTTTTTT
\begin{table}[!h]\small
            \caption{Hyperparameter values used in XG-BoT}
            \label{anomal-e_parameters}
            \centering
            \begin{tabular}{*2l }
                \toprule
                Hyperparameter & Values  \\
                \toprule
                No. Layers & $[3, 6, 9, 12, 15, 18, 21, 24]$ \\
                No. Hidden Channels & 64 (32 hidden channels for each GINs)\\
                No. Groups & 2\\
                Learning Rate & $1e^{-3}$\\
                Activation Func. & ReLU\\
                Optimiser & Adam\\ \hline
            \end{tabular}
                
        \label{lab:hypermanters}
        \end{table}
        
To demonstrate the computational efficiency of the XG-BoT model, we measured the training time for the optimal model and the Mean time to detect (MTTD). The performance results for the two benchmark datasets are shown in Table~\ref{lab:runtime}.

\section{Experimental Results}\label{results}

For the performance evaluation of the proposed XG-BoT, the evaluation metrics listed in Table~\ref{tab: metrics} were used, where $TP$, $TN$, $FP$ and $FN$ represent the number of True Positives, True Negatives, False Positives, and False Negatives, respectively.

Table~\ref{table:binary_comp} shows the performance evaluation results of the proposed \mbox{XG-BoT} model for automatic botnet detection, indicating the Precision, F1-Score, Recall, and FAR for the C2 and P2P datasets.
Overall, the XG-BoT model achieved state-of-the-art performance scores. As all the datasets are highly imbalanced, we did not use accuracy as a performance metric. The performance metrics show that XG-BoT achieves extremely low false alarm rates and very high detection rates, in both the C2 and P2P experiments.

%TTTTTTTTTTTTTTTTTTTTTTTTTTTTTTTTTTTTTTTTTTT
\begin{table}[!h]
\renewcommand{\arraystretch}{1.4}
\caption{Evaluation metrics utilised in this study.}
\centering
\begin{tabular}
{ c c} \hline
\textbf{Metric} & \textbf{Definition} \\ \hline 
\small{ Recall (Detection Rate)} &  $\frac{TP}{TP+FN}$ \\  \hline
\small{Precision} & $\frac{TP}{TP+FP}$ \\ \hline
\small{F1-Score} & $ 2 \times \frac{Recall\times Precision}{Recall + Precision}$ \\ \hline
\small{FAR (False Alarm Rate)} &   $\frac{FP}{FP+TN}$ \\ \hline
\end{tabular}
\label{tab: metrics}
\end{table}
%TTTTTTTTTTTTTTTTTTTTTTTTTTTTTTTTTTTTTTT
%
We then used detection rates and F1 and Recall scores to compare our proposed model with the state-of-the-art results, i.e., the best classification results shown in the literature. Table \ref{table:binary_comp} shows the corresponding results for the  \mbox{XG-BoT} classifier compared to the state-of-the-art results in terms of detection rates and F1-Score. As we can see in the table, for the C2 dataset experiment, XG-BoT achieved F1 and Recall scores of 99.52\% and 99.42\%, respectively. In the second P2P dataset experiment, XG-BoT achieved F1 and Recall scores of 99.47\% and 99.72\%, respectively. In regard to the F1-Score and Recall, XG-BoT outperformed all state-of-the-art approaches in both P2P and C2 experiments.

%TTTTTTTTTTTTTTTTTTTTTTTTTTTTTTTTTTTTTTTTTTTT
\begin{table*}[!h]
\centering
\caption{Performance of binary classification by XG-BoT compared with the state-of-art algorithms}
\begin{tabular}{ c  c c c c  c}
%  \hline
%  \multicolumn{4}{|c|}{Binary Classification Performance Comparison with State-of-Arts Algorithms} \\
 \hline
  \textbf{Method} & \textbf{Dataset} &  \textbf{Precision} & \textbf{Recall} & \textbf{F1} & \textbf{FAR} \\

\hline

\textbf{Proposed XG-BoT }  & C2 & 99.63\%   & 99.42\% & \textbf{99.52\%} & 0.01\%  \\
 GCN~\cite{zhou2020automating}  & C2  &  $-$  &  99.03\% & $-$ & 0.01\%  \\
 GCN~\cite{zhang2021practical}   & C2  &  $-$ & 96.40\% & 98.00\%  & $-$  \\
 XGBoost~\cite{zhang2021practical}  & C2  &  $-$ & 96.00\% & 11.80\%  & $-$  \\
   \hline

\textbf{Proposed XG-BoT }  & P2P   & 99.23\% & 99.72\% & \textbf{99.47\%} & 0.02\%  \\
 GCN~\cite{zhou2020automating}  & P2P  & $-$ & 99.51\% & $-$ & 0.01\% \\
 GCN~\cite{zhang2021practical}  & P2P  &  $-$  & 98.40\% & 98.91\%  & $-$   \\
 XGBoost~\cite{zhang2021practical}  & P2P  &  $-$  & 98.50\% & 10.20\% &  $-$   \\
 ABD-GN~\cite{carpenter2021detecting}  & P2P  & $-$  &  $-$  & 99.29\%  &  $0.01\%$  \\
 isirgn1~\cite{carpenter2021detecting}  & P2P  &  $-$   & $-$ & 97.85\% &  $0.02\%$  \\

  \hline

\end{tabular}

\label{table:binary_comp}
\end{table*}
%TTTTTTTTTTTTTTTTTTTTTTTTTTTTTTTTTTTTTTTTTTTT

\begin{figure}[!t]
    \centering
        \includegraphics[width=0.62\columnwidth]{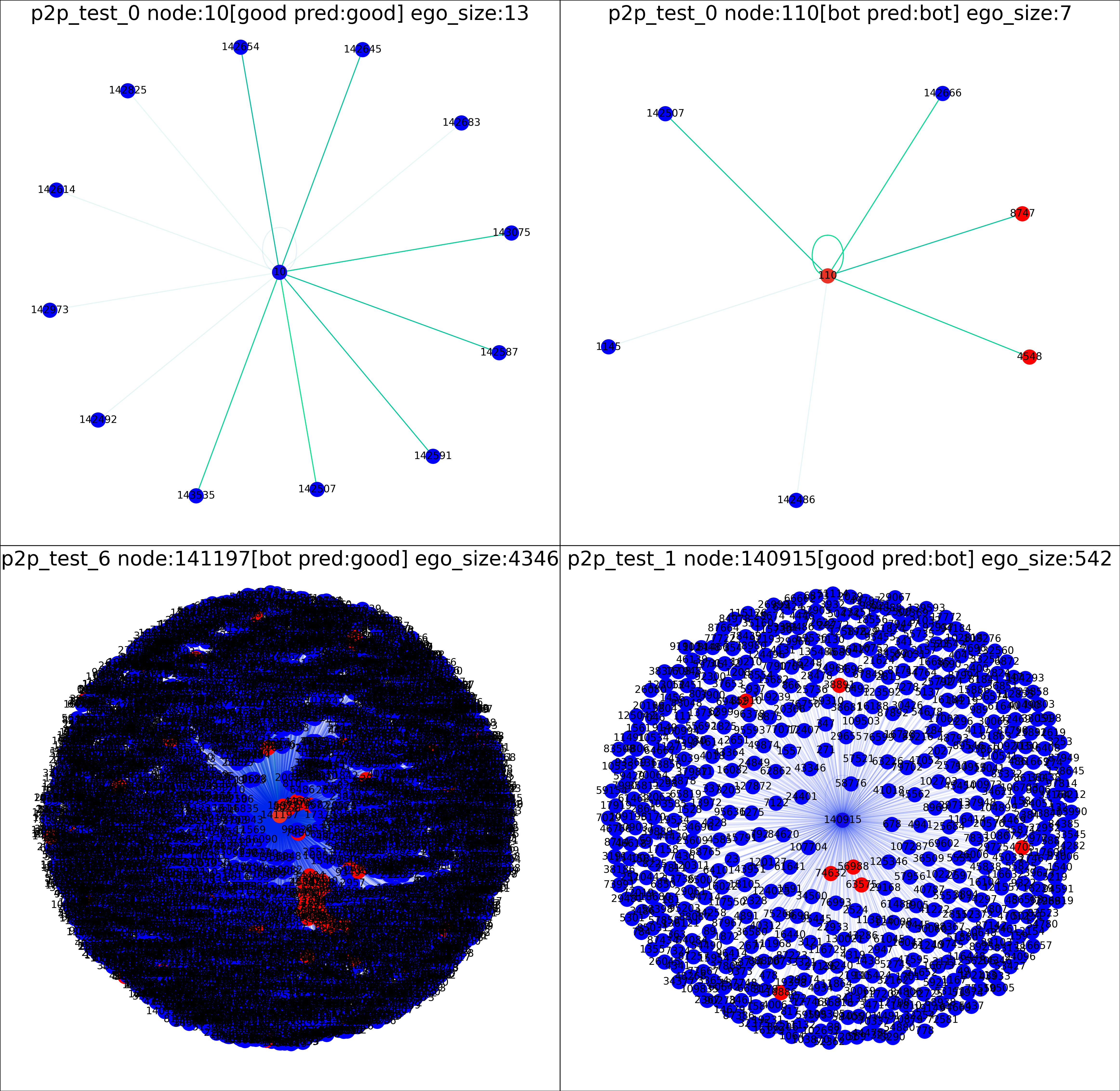}
     \caption{GNNExplainer for selected test nodes in the P2P dataset. Clockwise from top-left : TP (node 10), TN (node 110), FN (node 141197), and FP (node 140915) from the XG-BoT prediction, showing the (edge) structure information used for the classification decision. Bots are red, and non-bots are blue circles. The edge color indicates the magnitude of the effect. Edges shown in darker green indicate more significant contributions to the detection results. Edges shown in darker blue indicate less significant contributions to the detection results.}
        \label{fig:gnnexplainer_results}
\end{figure}

\section{Explainability for Automatic Network Forensics}\label{explainable_results}
While there is a huge interest in the explainability of deep learning model predictions \cite{kasanishi2021edge}, the adoption of GNN explainability involves some challenges \cite{sundararajan2017axiomatic}. In this section, we discuss the XG-BoT explainability methods, with a specific focus on automatic network forensics via subgraph visualization.

%In this paper, we also utilized GNNExplainer~\cite{ying2019gnnexplainer} for automatic network forensics by visualizing the subgraph. GNNExplainer~\cite{ying2019gnnexplainer} is an explainable graph algorithm that provides explanations for graph neural network prediction. 

\textbf{GNNExplainer} \cite{ying2019gnnexplainer}: This is an explainable graph algorithm that provides interpretable explanations for GNN predictions. Given an individual input graph, GNNExplainer emphasizes a subgraph structure and node-level features that are relevant to the prediction. In an explainable botnet node classification case the algorithm returns the most critically important host and network flow paths in an explainable botnet node classification case. GNNExplainer tries to maximize a mutual information objective function between the prediction of a graph neural network and the distribution of feasible subgraphs is maximized. The goal of GNNExplainer is to identify a subgraph $G_{\mathrm{S}} \subseteq G$ with associated features $X_{\mathrm{S}}=\left\{x_{j} \mid v_{j} \in G_{\mathrm{S}}\right\}$ that are relevant for explaining a target prediction via a mutual information measure MI, where $H$ is an entropy term. 

\begin{equation}
\begin{aligned}
&\max _{G_{\mathrm{S}}} \operatorname{MI}\left(Y,\left(G_{\mathrm{S}}, X_{\mathrm{S}}\right)\right)= 
&H(Y)-H\left(Y \mid G=G_{\mathrm{S}}, X=X_{\mathrm{S}}\right)
\end{aligned}
\end{equation}

The detection results of the XG-BoT can be explained by the contribution of "learnt" node features and the interconnections of nodes towards the suspected bot.
%In this paper, we used GNNExplainer \cite{ying2019gnnexplainer} for automatic network forensics. 
The explainable algorithm learns and returns a node feature and edge mask to explain the importance of corresponding nodes and edges that contributed to the final detection result. The masks can be learned using gradient descents by maximizing the mutual information between the subgraph and the final detection. Fig. \ref{fig:gnnexplainer_results}  shows the explainable results of the P2P botnet graph nodes with the normal/bot node samples. The edges in green (darker indicates more significant) are the most relevant path, which contributes to the detection of the centre targeted bot (shown in red) in Fig. \ref{fig:gnnexplainer_results} right and the targeted centre normal node (shown in blue) in Fig. \ref{fig:gnnexplainer_results} left. Overall, GNNExplainer correctly identifies corresponding neighbour nodes in the TP scenarios, which contributes to the detection results as it highlights highly correlated hosts and network flows.

%\textcolor{blue}{While GNNExplainer is able to find the subgraph and select features of the explained node as explanations, it mainly focuses on graph structures and not on finding useful features \cite{huang2022graphlime}. Local Interpretable Model Explanation (LIME,~\cite{ribeiro2016should}) models focus on finding features that explain the GNN's prediction. Specifically, LIME samples data with a perturbation-based sampling method to train a linear explanation model and then selects explanatory features according to the model's coefficients. GraphLIME~\cite{huang2022graphlime} extends LIME to use both graph features and structure of a node by sampling its N-hop neighbourhood and then computing the K most representative features that explain the prediction, using  Hilbert-Schmidt Independence Criterion Lasso (HSIC Lasso)~\cite{yamada2014high}.}

\begin{figure}[!t]
    \centering
        \includegraphics[width=0.62\columnwidth]{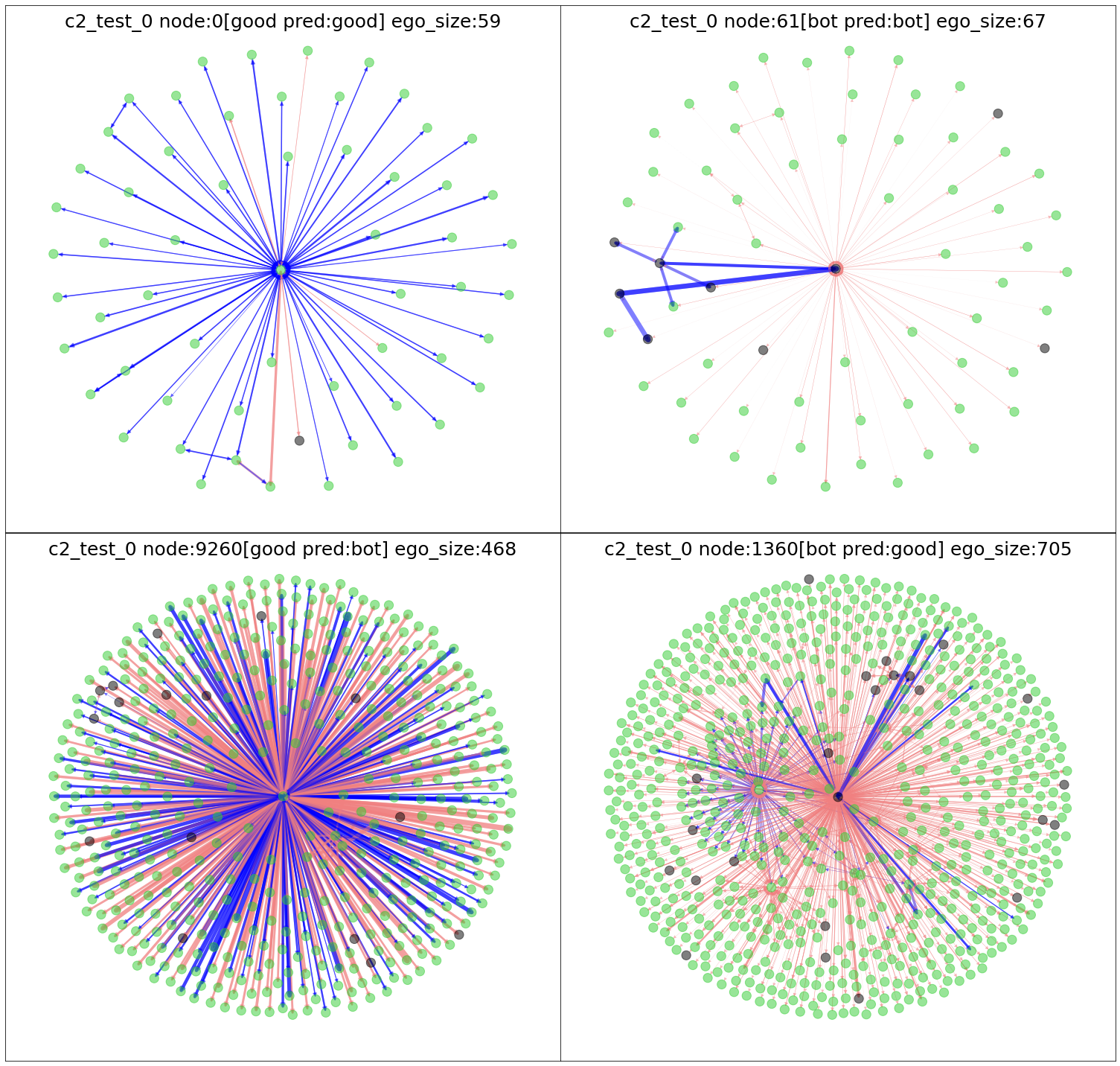}
    \caption{Saliency map for selected test nodes in the C2 dataset. Clockwise from top-left : TP (node 0), TN (node 61), FP (node 9260), and FN (node 1360) from the XG-BoT prediction showing the (edge) structure information used for the classification decision. Bots are black, and non-bots are presented as green circles. The edge width indicates the magnitude of the effect. Positive gradients are presented in red. Negative gradient changes are presented in blue.}
    \label{fig:SM}
\end{figure}

\textbf{Gradient-Based Saliency Maps} \cite{kasanishi2021edge}: These are the derivatives of the class probability $P_i$ to the input image $X$ \cite{Simonyan2014deep}, given by $\frac{dP_i}{dX}$. Essentially similar to backpropagation, they generate a heat map corresponding to each pixel's importance. This image domain CNN-based explainability method can be adopted to predict which edges are important for GNN decisions \cite{kasanishi2021edge}. Specifically, in the case of XG-BoT classification, a quantification is obtained for every neighbour node in a node's ego network with respect to a certain class prediction $P_i$.
We start with all connected edge weights set to one initially and calculate the gradient of the output with respect to the edge weights $w_{e_i}$. We use the absolute value of the gradient as the attribution value for each edge:
\begin{equation}
Attribution_{e_i}=  |\frac{\partial F(x)}{ \partial w_{e_i}} |,
\end{equation}
where x is the input and F(x) is the output of XG-BoT on input node x. Figure~\ref{fig:SM} shows the influence of the XG-BoT decision on a node from its ego network. In the FP and FN cases (bottom row), given the context, it is clear to see the reasoning for the XG-BoT classification decision.

\iffalse
In summary, GNN-based explainability is a challenging area \cite{sundararajan2017axiomatic} with few works available that explain GNN models \cite{kasanishi2021edge}. Sole reliance on visual inspection may be misleading \cite{adebayo2018sanity} and even unreasonable or impossible for structural graph data. Still, explanations may depend on the model architecture due to appropriately incorporated priors. In such situations, as in XG-BoT, the architecture can be sufficient for giving useful explanations based on explainable results, and the network administrator can use those results for automatic forensics. 
\fi

In summary, GNN-based explainability is challenging \cite{sundararajan2017axiomatic} with few works available to explain GNN models \cite{kasanishi2021edge}. Unlike purely visual inspections, which can be misleading \cite{adebayo2018sanity}, identifying influential neighbours via salience or gradients is simple and efficient and is recommended for XG-BoT-based detection. For datasets where features are not present, using perturbation-based methods to monitor changes
in the predictions by perturbing different input features is not effective. However, XG-BoT can be used with Integrated Gradients \cite{sundararajan2017axiomatic} to assign an importance score based on an approximation of the integral of the gradient or Clustering-based approaches \cite{kulatilleke2022scgc} to provide explainable results. The network administrator can use those results for automatic forensics.

\section{Conclusion and Future Work}\label{summary}
This paper proposes a novel, explainable GNN-Based botnet detection system. We first present XG-BoT, which uses the grouped reversible residual connection and a graph isomorphism network to perform botnet detection. Then GNNExplainer and saliency map was applied by highlighting specious network flows and botnet nodes. Given the experimental results, two benchmark datasets indicate that our approach can outperform the state-of-the-art approaches in terms of F1-Score and detection rate. Furthermore, identifying the suspicious network flows and botnet nodes can facilitate understanding botnet patterns for automatic network forensics. In the future, we plan to apply edge-based graph encoders, such as E-GraphSAGE \cite{9789878}, with XG-BoT to consider network communication flow features, such as the number of packets required for deep-wise edge-based botnet detection.

\section*{Declaration of Competing Interest}
The authors declare that they have no known competing financial interests or personal relationships that could have appeared to influence the work reported in this paper.

\bibliography{main}

\end{sloppypar}
\end{document}